\documentclass[twocolumn,showpacs,preprintnumbers,amsmath,amssymb]{revtex4}

\usepackage{graphicx}
\usepackage{dcolumn,subfigure,enumitem}
\usepackage{bm}

\begin{document}

\title{Astrophysical Quantum Matter: Spinless charged particles on a magnetic 
dipole sphere}

\author{Jeff Murugan}
\author{Jonathan P. Shock}
\author{Ruach Pillay Slayen}

\affiliation{The Laboratory for Quantum Gravity \& Strings and, Department of Mathematics and Applied Mathematics, Univeristy of Cape Town,Private Bag, Rondebosch 7700, South Africa}
\date{\today}

\begin{abstract}
   We consider the quantum mechanics of a spinless charged particle on a 2-dimensional sphere. 
   When threaded with a 
   magnetic monopole field, this is the well-known Haldane sphere that furnishes a translationally-invariant,
   incompressible quantum fluid state of a gas of electrons confined to the sphere. 
   This letter presents the results of a novel 
   variant of the Haldane solution where the monopole field is replaced by that of a dipole. 
   We argue that this system is relevant to the physics on the surface of compact astrophysical objects 
   like neutron stars.
\end{abstract}


\maketitle

\section{Introduction}
Condensed matter phenomena provide some of the most spectacular vistas into the Quantum Realm 
\cite{Wright:2015}. 
The discoveries of the quantum Hall and fractional quantum Hall effects, topological quantum matter and 
the electronic properties of graphene are some of the most celebrated of a wealth of new phenomena discovered over the past two decades. These discoveries have shaped this rapidly developing field 
as more and  more of the quantum nature of matter is uncovered. Much of 
this was discovered by studying the quantum dynamics of charged particles in strong magnetic fields.  \\

In attempting to realize a translationally invariant incompressible  
(Laughlin) quantum fluid state in the early 1980's, Haldane studied a 2-dimensional electron gas on a 
sphere of radius $R$. This `{\it Haldane sphere}' was then threaded by a constant magnetic field, 
$\bm{B} = \left(\hbar Q/eR^{2}\right)\widehat{\bm{r}}$, produced by a monopole at its center 
\cite{Haldane:1983xm}. In contrast to the more studied planar geometry, Haldane's 
configuration produces quantum states for which the Landau levels have finite degeneracy. Consequently, the notion of 
a filled Landau level can be unambiguously defined, without an external confining potential. Without boundaries, 
and the accompanying subtleties of edge modes, compact geometries provide a good way to probe the bulk physics 
of the Landau problem \cite{Jain:2007}. The years following Haldane's work has seen a slew 
of investigations of the properties of charged particles on various compact (and non-compact) spaces. 
These include tori \cite{Haldane:1985eda}, cylinders \cite{Bellucci:2010} and
higher-genus Riemann surfaces \cite{Wen:1990zza}, all of which point to a remarkably rich  
mathematical structure in addition to the complex phenomenology of the quantum Hall effect.\\

Phenomenology, specifically of the tabletop variety, is the chief reason that the quantum Hall
effect, and its fractional cousin have generated so much excitement since their discoveries. However, such 
tabletop experiments are constrained by the intensity of the magnetic fields that can be produced on
Earth\footnote{While $\sim 10^{14}$T magnetic fields are achievable in heavy-ion collisions at, for example 
the LHC, these are out of equilibrium and too short-lived to be relevant for our purposes.}. 
Consequently, if we want to learn more about quantum matter interacting with stable magnetic fields stronger than the 
$\sim \mathcal{O}(100)$T fields achievable in terrestrial laboratories \cite{Helmholtz}, it makes sense 
to turn our attention skyward. With magnetic fields that range in strength from $10^{4} - 10^{11}$T, neutron 
stars, including their more powerfully magnetic versions, magnetars, exhibit some of the most intense magnetic fields
in the known Universe. Moreover, in this age of precision multi-messenger observations, we have 
access to volumes of astrophysical data that offer an unprecedented glimpse into the physics of these 
gigantic generators. While much effort is directed toward probing the interior of neutron stars, there is also
a considerable amount to be learnt from their surface (see, for example, \cite{Heyl:2018kah} for some
recent developments in neutron star atmospheric physics). The essential 
motivation for this letter was piqued by the question: {\it Are there astrophysical signatures of topological quantum 
matter}?\\
 
One crucial difference between the physics of a 2-dimensional gas of electrons on the surface of a neutron star 
and the Haldane sphere is that, in the former, the magnetic field is primarily {\it dipole} in nature\footnote{Although 
it must be pointed out that it is likely not {\it pure} dipole. Like the sun, there are multipole moments that decay 
rapidly with radius. We thank Bryan Gaensler for reminding us of this.}. Consequently, 
the net magnetic flux ($2Q$ in the Haldane problem) is zero. Nevertheless, in the large $R$ limit and with the 
field lines sufficiently focused at the poles, charged particles in the polar neighbourhoods find themselves trapped 
in the cyclotron motion of the planar Hall effect. Of course, the study of charged classical particles in 
dipole fields has a long history in the geophysical context of cosmic rays and auroral phenomena 
\cite{Stormer:1907}. The equations of motion for this St\"ormer problem are second order, coupled and 
nonlinear. Unsurprisingly, for general initial data, they do not possess any simple analytic solutions. Still, at 
sufficiently low energies the dynamical system exhibits a rich class of orbits trapped about guiding field 
lines\cite{Dragt:1965}. While we will not require anything so elaborate, this analysis will serve as a useful 
starting point.

\section{Classical dynamics in a dipole field}
To set up the system, we replace the physically unrealistic magnetic monopole at the center of the Haldane 
sphere with {\it two} monopoles with charges $+b$ and $-b$ separated by a distance $l$ and 
aligned along the $\widehat{\bm{z}}$-direction. In the limit that $l\to 0$ with $bl$ fixed, the resulting 
magnetic field 
\begin{eqnarray}\label{Bfield}
   \bm{B} = \frac{|\bm{\mu}|}{r^{3}}\left(2\cos\theta\,\widehat{\bm{r}} 
   + \sin\theta\, \widehat{\bm{\theta}}\right)\,,
\end{eqnarray}
is identical to that produced by a current loop enclosing an area $A_{\rm loop}$ and oriented in the $xy$-plane. 
In either case the magnetic moment is aligned in the positive $\widehat{\bm{z}}$-direction with 
$|\bm{\mu}| = IA_{\rm loop} = bl$. Associated to this dipole magnetic field is the  vector potential
\begin{eqnarray}\label{pot}
   \bm{A} = \frac{1}{r^{2}}\bm{\mu}\times\widehat{\bm{r}} = \frac{|\bm{\mu}|}{r^{2}}\sin\theta\,
   \widehat{\bm{\phi}}\,.
\end{eqnarray}
A particle of charge $q$ and mass $m$, confined to move on a sphere of radius $R$ concentric with the center of 
the dipole, does so with Hamiltonian (in units of $c=1$)
\begin{eqnarray*}
   H = \frac{1}{2m}\bm{\Pi}^{2} = \frac{1}{2m}(\bm{p} -q\bm{A})^{2}\,.
\end{eqnarray*}
The resulting equations of motion
\begin{eqnarray*}
   \ddot{\theta} &=& \frac{1}{2}\sin2\theta\left(\dot{\phi} + \frac{2q|\bm{\mu}|}{mR^{3}}\right)
   \dot{\phi}\,,\\
   \ddot{\phi} &=&-2\cot\theta\left(\dot{\phi} + \frac{q|\bm{\mu}|}{mR^{3}}\right)\dot{\theta}\,, 
\end{eqnarray*}
can be numerically solved for a rich set of trajectories. Depending on the initial conditions, at least some of these are 
closed trapped orbits that resemble the Haldane sphere case (Fig.2.(a)). Others  (Fig.2.(b)) are more elaborate 
and closer to the classical orbits of a planar particle in a spatially-varying transverse magnetic field 
\cite{Grosse:1993np}.

\begin{figure}
   \includegraphics[height=3cm,width=8.5cm]{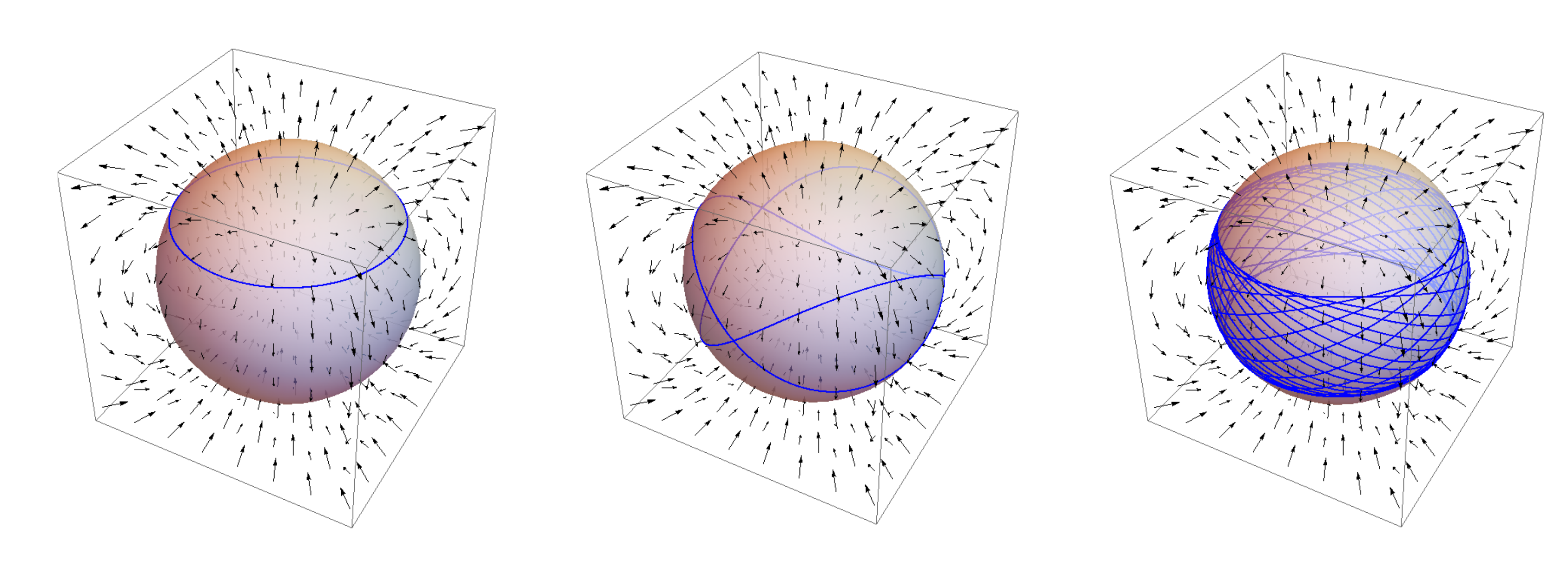}
   \caption{\label{classical} Classical particle orbits on the dipole sphere.}
\end{figure}
\section{Quantum Mechanics on the dipole sphere}
Now let's study the {\it quantum mechanics} of charged particles constrained to the surface of the sphere surounding a magnetic dipole at its centre. In the monopole problem where the charged particle moves on a 2-dimensional 
surface interacting with a uniform and everwhere-perpendicular magnetic field, the problem shares many similarities 
with the planar Landau problem. Indeed, in the large $R$ limit, it is expected that the two problems converge. 
For the dipole, this is no longer the case. Even if the particle is constrained by its initial conditions to move on
orbits of latitudes (see Fig.2(a)), the magnetic field is no longer uniform, nor perpendicular. Intuitively, we would 
expect that in the large radius limit, when the sphere is nearly flat, and sufficiently close to the north pole the 
system reduces to the planar Landau problem but in an {\it inhomogeneous} magnetic field with rotational symmetry in the $\phi$ direction. On the other hand, orbits that transit the equator,
as in Fig. 2(b), will not. As a first step toward answering the question posed in the introduction, here we will make some of these notions more precise. Our treatment will follow that in \cite{Fano:1986zz} for the monopole sphere, suitably adapted to the dipole. \\

We start by writing the Hamiltonian in the form $H = (\hbar^{2}/mR^{2})\bm{\Lambda}\cdot \bm{\Lambda}$,
where the angular momentum operator
\begin{eqnarray*}
   \bm{\Lambda} = -i\left(\widehat{\bm{\phi}}\frac{\partial}{\partial\theta} 
   - \widehat{\bm{\theta}}\frac{1}{\sin\theta}\frac{\partial}{\partial\phi}\right) - 
   q|\bm{\mu}|\sin\theta\,\widehat{\bm{\theta}}\,.
\end{eqnarray*}
Consequently, the single-particle Schr\"odinger equation reads
\begin{eqnarray*}
   E\Psi = \frac{\hbar^{2}}{2mR^{2}}\!\!\left[-\frac{1}{\sin\theta}\frac{\partial}{\partial\theta}\sin\theta
   \frac{\partial}{\partial\theta} + \left(\frac{i}{\sin\theta} \frac{\partial}{\partial\phi}
   - Q\sin\theta\right)^{2}\right]\!\!\Psi\,,
\end{eqnarray*}
where, to facilitate comparison with the monopole sphere, we have defined $Q\equiv q|\bm{\mu}|/\hbar R$. 
We introduce a separable ansatz $\Psi(\theta,\phi) = e^{im\phi}\psi(\theta)$, which puts the eigenvalue 
problem in the form
\begin{eqnarray*}
   \widetilde{E}\psi &=& -\frac{1}{\sin\theta}\frac{\partial}{\partial\theta}\left(\sin\theta
   \frac{\partial\psi}{\partial\theta}\right) +\frac{\left(m+Q\sin^{2}\theta\right)^{2}}
   {\sin^{2}\theta}\psi\,,
\end{eqnarray*}
where, to avoid confusion between the mass of the 
particle and the quantum number labelling the eigenfunction, we define 
$\widetilde{E} \equiv 2mR^{2}E/\hbar^{2}$. Now we set 
$z = \cos\theta$ and write the differential equation in the algebraic form,
\begin{eqnarray}\label{sph}
    \frac{d}{dz}\left[\left(1-z^{2}\right)\frac{d}{dz}\right]\psi 
    + \left[\lambda_{m,l}+Q^{2}z^{2}-\frac{m^{2}}{(1-z^{2})}\right]\psi=0,\,\,
\end{eqnarray}
with $\lambda_{m,l} \equiv \widetilde{E}- 2mQ - Q^{2}$.  This is the {\it angular 
oblate spheroidal equation} \cite{Morse:1953}, whose solutions are the angular 
oblate spheroidal wavefunctions, $S_{m,l}(Q,z)$, labelled by the spheroidal harmonic index $l$ with 
$l-|m| = 0,1,2,\ldots$ The eigenvalues $\lambda_{m,l}$ are fixed by the requirements
that the wavefunction remain finite at $z=\pm1$, are real-valued and satisfy the conjugation relation 
$\lambda_{l,m} = \lambda_{l,-m}$. 
In hindsight, the emergence of spheroidal  symmetry should not be 
surprising since the presence of the dipole magnetic field breaks the rotational symmetry of the Haldane problem 
to that of a flattened (oblate) sphere. In order to write down the normalised single-particle eigenstates, 
we now elaborate on these solutions.

\section{Single-particle states}
The angular spheroidal wave equation \eqref{sph} has two regular singular points at $z= 1$ and $z=-1$,
corresponding to the north and south poles of the sphere respectively. Solutions of 
this equation can be expressed as a sum,
\begin{eqnarray}\label{sphfun}
    S_{l,m}(Q,z) =  (1-z^{2})^{m/2}\sum_{n}d_{n}^{l,m}(Q)\frac{d^{m}
    P_{n}(z)}{dz^{m}}\,,
\end{eqnarray}
if we use the fact that in the interval $z\in[-1,+1]$, the associated Legendre functions can be expressed as derivatives
of Legendre polynomials (of the first kind),
$P_{n}^{m}(z) = (1-z^{2})^{m/2}d^{m}P_{n}(z)/dz^{m}$. The coefficient functions satisfy a three-term recursion relation 
\cite{Morse:1953},
 \begin{eqnarray*}
   \alpha_{n}d_{n+2} + \left(\beta_{n}-\lambda_{l,m}\right)d_{n} + \gamma_{n}d_{n-2} = 0\,,
\end{eqnarray*}
where the coefficients $\alpha_{n}, \beta_{n}$ and $\gamma_{n}$ are not relevant for the purposes of 
this discussion. The recursion relation may be solved by, for example, the method of 
continued fractions. In general, solutions that are finite at $z=\pm 1$ 
will diverge at $z=\mp 1$, but for a discrete set of values of the eigenvalues $\lambda_{l,m}$, the series will 
converge to solutions that are finite at both poles. In fact, there are {\it two} sets of finite solutions, one for even $n$
and one for odd, so that the sum in \eqref{sphfun} runs separately over even $n$ with $l=m,m+2,\ldots$, and 
odd $n$ for which $l=m+1,m+3,\ldots$, and with $\lambda_{l,m}<\lambda_{l+1,m}$.  Of particular interest to
 us will be the fact that:
 \begin{itemize}[leftmargin=*]
  \item For a given value of $m$, the lowest value of the eigenvalue is that for which $l=|m|$. Also, for fixed $m$, the 
    corresponding set of eigenfunctions with different $l$ values are mutually orthogonal. Consequently the full
    wavefunctions satisfy $\langle \Psi_{Q,l,m}(\theta,\phi)|\Psi_{Q,l',m'}(\theta,\phi)\rangle = 
    \delta_{ll'}\delta_{mm'}$.
 \item In the $Q\to 0$ limit, the equation for $S(z)$ reduces to the equation for a single spherical harmonic 
    $P^{m}_{l}(z)$ with the corresponding eigenvalues $\lambda_{l,m}= l(l+1)$, as expected for a free particle
    confined to the surface of a sphere. 
 \item There are a number of normalisation schemes for the angular oblate functions. In the Stratton-Morse 
   scheme which will be most convenient for our purposes, $S$ can be normalised by imposing that, 
   near $z=1$, it behaves like $P^{m}_{l}$ for all values of $Q$. 
   This in turn requires that the expansion coefficients satisfy
    \begin{eqnarray*}
       \widetilde{\sum_{n}}\,\frac{(n+2m)!}{n!}d_{n}^{l,m}(Q) = \frac{(l+m)!}{(l-m)!}\,.
    \end{eqnarray*}
    The tilde over the summation sign is an instruction to include only even values of $n$ if $(l-m)$ is even 
    and only odd values of $n$
    if $(l-m)$ is odd. With this, the normalisation constants
    \begin{eqnarray*}
    &&(\mathcal{N}_{l,m})^{-1} =\int_{-1}^{1}\left(S_{l,m}(Q,z)\right)^{2}\,dz\\
    \quad\quad&=& \widetilde{\sum_{n}}\,\left(d_{n}^{l,m}(Q)\right)^{2}\left(\frac{2}{2n+2m+1}\right)
    \left(\frac{(n+2m)!}{n!}\right).
    \end{eqnarray*}
\end{itemize}
Drawing this all together, the normalised single-particle eigenstates are given by
\begin{eqnarray*}\label{single}
   \Psi_{Q,l,m}(\theta,\phi) &=& \mathcal{N}_{l,m}\,e^{im\phi}(1-z^{2})^{m/2}
    \widetilde{\sum_{n}} d_{n}^{l,m}(Q)\frac{d^{m}
    P_{n}(z)}{dz^{m}}\,,
 \end{eqnarray*}
with $z=\cos\theta$ and corresponding energy eigenvalues 
\begin{eqnarray*}
   \widetilde{E}_{Q,l,m} = \lambda_{l,m}+2mQ+Q^{2}\,.
\end{eqnarray*}
Even for the very restricted class of orbits that we have considered, this system clearly posseses a rich set of
solutions. While we will explore these in greater detail elsewhere \cite{Murugan:2018}, some intuition for the physics
can be built by studying the solutions in some limiting cases.\\

\begin{itemize}[leftmargin=*]
 \item {\it Near the north pole}, the polar angle $\theta\approx 0$ and $z\to 1$. In the Stratton-Morse normalisation, the behaviour of 
$S_{m,l}$ remains close to the associated Legendre function for all values of $Q$, and so, expanding in a 
power series in $(1-z^{2})$,
\begin{eqnarray*}
   \quad\qquad\Psi_{Q,l,m} = \mathcal{N}_{l,m}\,e^{im\phi}(1-z^{2})^{m/2}\sum_{k=0}^{\infty}
   c_{2k}^{l,m}(1-z^{2})^{k}\,,
\end{eqnarray*}
when $l-m$ is even, and where the coefficients $c_{2k}^{l,m}$ are expressed as a sum over the $d_{2k}^{l,m}$.
A similar expression holds when $l-m$ is odd.
\item {\it The weak-field limit} is what we will call $Q\ll 1$. 
For fixed value of the magnetic dipole moment, this is obtained by taking the large-$R$ limit. 
Using the known power-series expansions for $S_{l,m}$, 
$\lambda_{l,m}$ and the $\mathcal{N}_{l,m}$ we can write down the corresponding expansions for 
the eigenstates and eigenenergies. For example, for $l=m=0$ and to $\mathcal{O}(Q^{4})$, 
\begin{eqnarray*}
   \quad\qquad\Psi_{Q,0,0} &=& \frac{1}{2}P_{0}(z) + \frac{1}{18}\left(P_{0}(z)+
   P_{2}(z)\right)Q^{2}\\
   &+&\frac{-364P_{0}(z) + 225P_{2}(z) + 27P_{4}(z)}{28350}Q^{4}\,,\\
   E_{Q,0,0} &=& \frac{\hbar^{2}}{2mR^{2}}\left(\frac{2}{3}Q^{2} - \frac{2}{135}Q^{4}\right)\,.
\end{eqnarray*}
\item {\it The strong-field limit}, $Q\gg1$ is relevant in the context of  the surface 
physics of a neutron star where the magnetic dipole moment $|\bm{\mu}|\sim 10^{35}$ T\,m$^{-3}$ 
and $R\sim 10^{4}$m. 
In this limit, $S_{l,m}(\theta,\phi)$ can be expanded in a series of Laguerre 
polynomials, giving a large-$Q$ approximation to the single-particle eigenstate. In particular 
(\cite{Berti:2005gp,Hod:2015cqa}), for $Q^{2}\gg m^{2}$, the spheroidal eigenvalues 
$\lambda_{l,m} = -Q^{2} + 2[l+1 - 
\mathrm{mod}(l-m, 2)]Q + \mathcal{O}(1)$, result in the approximately (for finite $Q$) {\it doubly-degenerate} energy 
spectrum plotted (for various $m\geq 0$) in Fig.2. Note the (approximate) Landau-level structure. 
For any fixed $m$ in this range, as $Q$ is increased the energies of 
states with pairwise adjacent $l$ values converge. This convergence happens faster for smaller $l$ values. 
\end{itemize}

\begin{figure}
   \includegraphics[height=4cm,width=5.5cm]{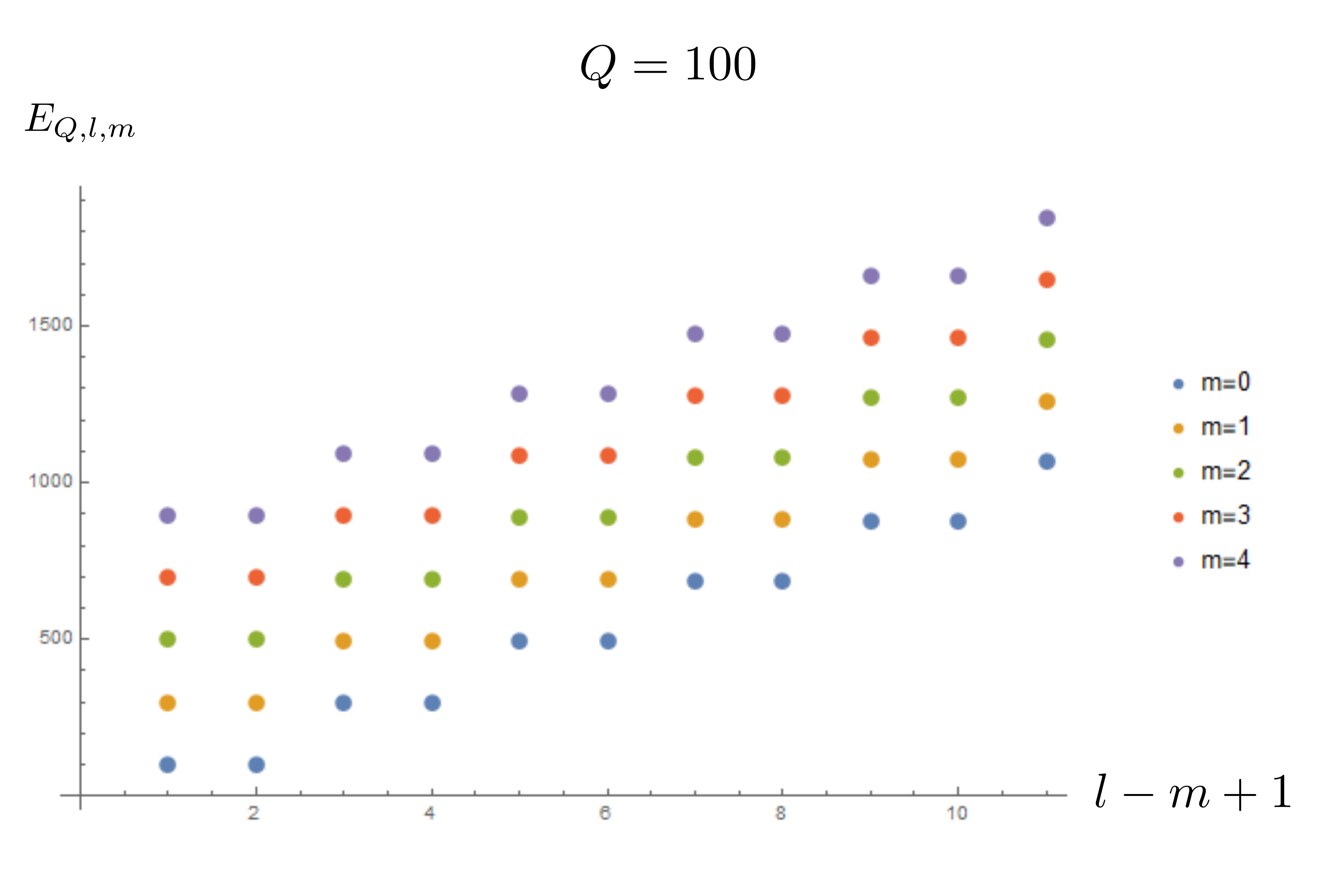}
   \caption{\label{spectrum} The energy spectrum in the large $Q$ limit.}
\end{figure}

\section{Observations and Discussion}
The Haldane model of a 2-dimensional gas of charged particles confined to a sphere in a backgound 
magnetic monopole field is the prototype of a clean system exhibiting the bulk physics of incompressible quantum 
fluids. In order to initiate a program into the possibility of observing extreme quantum phenomena in the 
astrophysical setting of neutron stars and pulsars,  in this letter we have extended the Haldane model to spinless
charged particles moving in a {\it dipole} field produced by two oppositely charged monopoles. 
While many of the computational techniques employed in the monopole case carry through for the dipole, 
the physics is markedly different. \\

\begin{figure}
   \includegraphics[height=3cm,width=7cm]{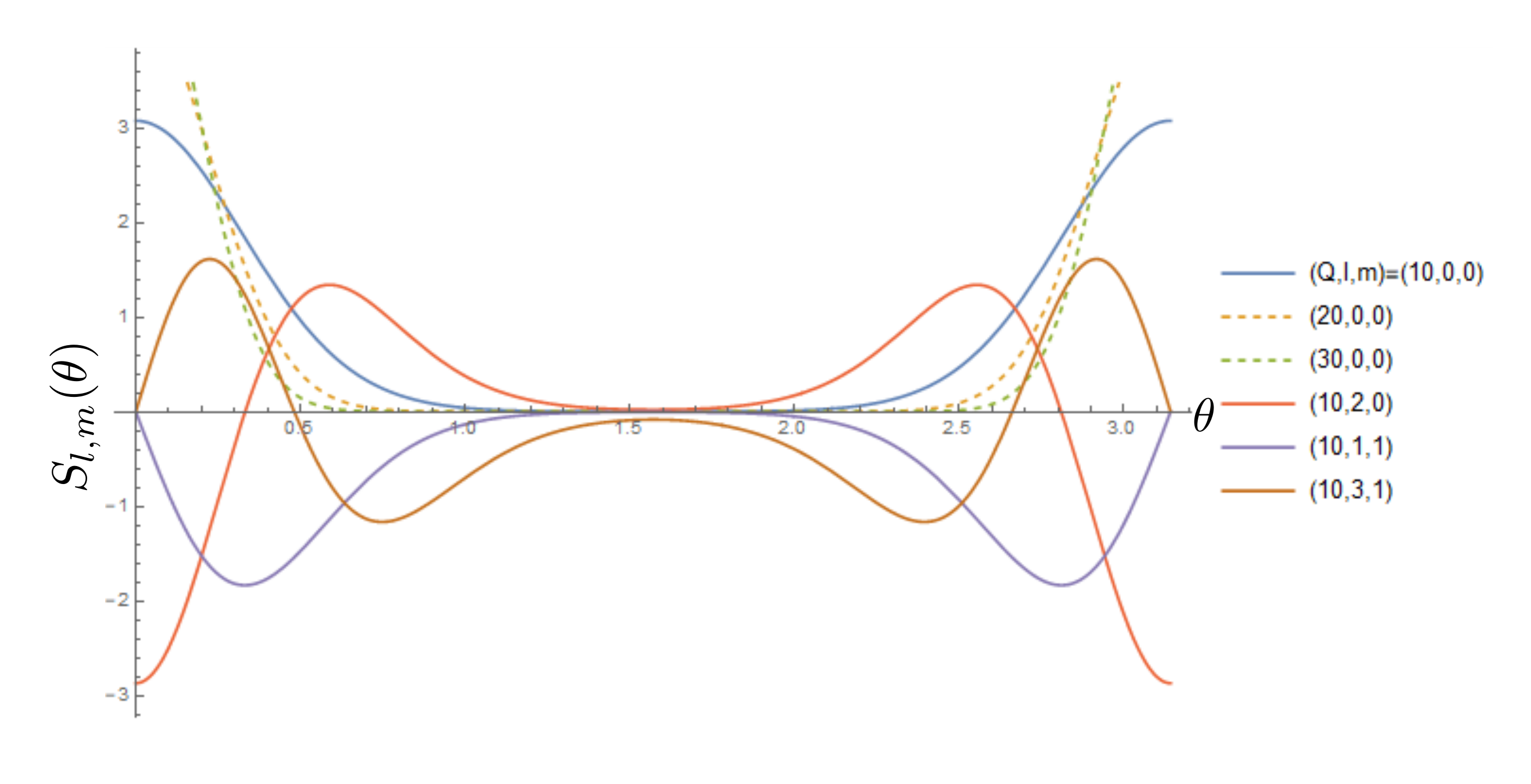}
   \caption{\label{wavefns2} $\Psi_{Q,l,m}(\theta,0)$ for several low-lying states.}
\end{figure}

There are several points of departure:
Because the dipole aligned along the $\hat{\bm{z}}$-axis breaks the $O(3)$ symmetry to a 
$U(1)$,  not all particle orbits are equivalent. In this letter we have focussed on the set of 
closed orbits parallel to the equatorial plane and centered on the north pole, along which the charged 
particle moves with 
cyclotron frequency $\omega_{c}$. For this set of orbits, the problem more closely resembles particle 
motion in an inhomogeneous magnetic field. We find that for a given $Q$, the $l=m=0$ state is 
the lowest energy state, and is localised around the poles. The stronger the magnetic field, the more 
pronounced this localisation becomes. Unlike the monopole case, the net magnetic flux through the sphere vanishes 
for the dipole making the problem globally topologically trivial. Nevertheless, it exhibits a remarkably 
rich structure. For sufficiently large magnetic fields, the energy levels display a Landau-like structure. 
However, unlike on the Haldane sphere, each level is infinitely degenerate. This is because, in the large
$Q$ limit, the energy, $\widetilde{E}_{Q,l,m} = 2[l+m+1- \mathrm{mod}(l+m,2)]Q + \mathcal{O}(1)$, 
depends only 
on the combination $l+m$. Consequently, all states with the same value of $l+m$ (with $l\geq|m|$) will have the 
same energy.\\

Undoubtably, we have only scratched the surface of this class of problems and are left with many more questions 
than answers. Given that we can't probe neutron stars the same way as, for example, a prepared laboratory 
sample the most pressing of these must be: {\it how do we extract this physics from neutron star observations}? 
Quantum Hall systems and their variants
are, by now, a staple of contemporary condensed matter physics yet, perhaps with the exception of high energy 
theory, remain largely unknown (at least in their details) outside the community. On the other hand,  
strongly magnetic, compact astrophysical objects, like neutron stars are the new 
``wild west of physics" and while much effort has been devoted to understanding the processes in the {\it interior} of 
such objects, comparatively little study has gone into studying surface phenomena, at least from the perspective 
of quantum matter. Rapidly increasing
sensitivities in astrophysical observations and forthcoming experiments focused 
on pulsars \cite{Kramer:2015}, promise an unprecedented opportunity to study quantum matter 
under extreme conditions. Our intent in this letter was as much to study the novel physics of particles in dipolar 
fields, as it was to bring to the attention of both the condensed matter and astrophysics communities, 
a potentially new and remarkable class of problem.  We hope that, if nothing else, it will stimulate further ideas in 
this direction.
\section{Acknowledgements}
We would like to thank Bryan Gaensler for very useful comments on the manuscript. 
JM is supported by the NRF of South Africa under grant 
CSUR 114599. RPS is supported by a graduate fellowship from the National Institute for Theoretical Physics.

\end{document}